\documentclass[times, tighten, twocolumn, twocolappendix]{aastex631}

\usepackage{savesym}
\savesymbol{tablenum}
\usepackage{siunitx}
\restoresymbol{SIX}{tablenum}

\usepackage{amsmath}
\usepackage{ulem}

\submitjournal{ApJL}

\begin{document}

\title{Size Dependence of the Bouncing Barrier in Protoplanetary Dust Growth}

\correspondingauthor{Sota Arakawa}
\email{arakawas@jamstec.go.jp}

\author[0000-0003-0947-9962]{Sota Arakawa}
\affiliation{Japan Agency for Marine-Earth Science and Technology, 3173-25 Showa-machi, Kanazawa-ku, Yokohama, 236-0001, Japan}

\author[0000-0002-1886-0880]{Satoshi Okuzumi}
\affiliation{Department of Earth and Planetary Sciences, Tokyo Institute of Technology, 2-12-1 Ookayama, Meguro, Tokyo, 152-8550, Japan}

\author[0000-0003-1844-5107]{Misako Tatsuuma}
\affiliation{Department of Earth and Planetary Sciences, Tokyo Institute of Technology, 2-12-1 Ookayama, Meguro, Tokyo, 152-8550, Japan}

\author[0000-0001-9659-658X]{Hidekazu Tanaka}
\affiliation{Astronomical Institute, Graduate School of Science, Tohoku University, 6-3 Aramaki, Aoba-ku, Sendai, 980-8578, Japan}

\author[0000-0002-5486-7828]{Eiichiro Kokubo}
\affiliation{National Astronomical Observatory of Japan, 2-21-1, Osawa, Mitaka, Tokyo, 181-8588, Japan}

\author[0000-0001-9163-9296]{Daisuke Nishiura}
\affiliation{Japan Agency for Marine-Earth Science and Technology, 3173-25 Showa-machi, Kanazawa-ku, Yokohama, 236-0001, Japan}

\author[0000-0002-4334-0879]{Mikito Furuichi}
\affiliation{Japan Agency for Marine-Earth Science and Technology, 3173-25 Showa-machi, Kanazawa-ku, Yokohama, 236-0001, Japan}

\author[0000-0003-3924-6174]{Taishi Nakamoto}
\affiliation{Department of Earth and Planetary Sciences, Tokyo Institute of Technology, 2-12-1 Ookayama, Meguro, Tokyo, 152-8550, Japan}



\begin{abstract}

Understanding the collisional behavior of dust aggregates is essential in the context of planet formation.
It is known that low-velocity collisions of dust aggregates result in bouncing rather than sticking when the filling factor of colliding dust aggregates is higher than a threshold value.
However, a large discrepancy between numerical and experimental results on the threshold filling factor was reported so far.
In this study, we perform numerical simulations using soft-sphere discrete element methods and demonstrate that the sticking probability decreases with increasing aggregates radius.
Our results suggest that the large discrepancy in the threshold filling factor may reflect the difference in the size of dust aggregates in earlier numerical simulations and laboratory experiments.

\end{abstract}



\section{Introduction}
\label{sec:intro}

In protoplanetary disks, the pairwise collisional growth of dust aggregates consisting of submicron-sized grains triggers planet formation.
Understanding the outcomes of collisions between dust aggregates is therefore essential, and the collisional behavior of dust aggregates has been studied by using both laboratory experiments \citep[e.g.,][]{1998Icar..132..125W, 2000Icar..143..138B, 2008ARA&A..46...21B, 2010A&A...513A..56G, 2012ApJ...758...35S, 2022MNRAS.509.5641S, 2017ApJ...836...94W, 2021ApJ...923..134F} and numerical simulations \citep[e.g.,][]{1997ApJ...480..647D, 2009ApJ...702.1490W, 2013A&A...559A..62W, 2016A&A...589A..30G, 2021ApJ...915...22H, 2023ApJ...944...38H, 2021MNRAS.503.1717P, 2022ApJ...933..144A, 2022ApJ...939..100A, 2023A&A...670L..21A, 2023ApJ...945...68T}.
It is widely accepted that the collisional outcomes are classified into three categories: sticking, bouncing, and fragmentation \citep[e.g.,][]{2010A&A...513A..56G}.

Laboratory experiments have reported that low-velocity collisions of dust aggregates result in bouncing rather than sticking under certain conditions \citep[e.g.,][]{2008ApJ...675..764L, 2011ApJ...736...34B, 2012A&A...542A..80J, 2012ApJ...758...35S, 2022MNRAS.509.5641S, 2012Icar..218..688W, 2013Icar..225...75K, 2014ApJ...783..111K, 2016A&A...593A...3B, 2017A&A...603A..66B, 2019A&A...631A..35B}.
The bouncing behavior has also been reported in earlier numerical simulations using soft-sphere discrete element methods \citep[e.g.,][]{2011ApJ...737...36W, 2012ApJ...758...35S, 2013A&A...551A..65S, OSINSKY2022127785}.
As bounding could limit dust growth in protoplanetary disks \citep[e.g.,][]{2010A&A...513A..57Z, 2018A&A...611A..18L}, the threshold condition for collisional sticking/bouncing is the key to understanding the pathway of planet formation.

The outcomes of low-velocity collisions between dust aggregates depend on many parameters, including the filling factor of the aggregates, the radius and composition of the constituent particles, and the collision velocity \citep[e.g.,][]{2010A&A...513A..56G, 2012A&A...542A..80J, 2022MNRAS.509.5641S}.
The degree of sintering also affects the collisional sticking/bouncing \citep[e.g.,][]{1999A&A...347..720S, 2012Icar..221..310S, 2017ApJ...841...36S, 2021ApJ...911..114S}.

The threshold filling factor for collisional sticking/bouncing have been investigated in several studies.
\citet{2011ApJ...737...36W} performed numerical simulations of low-velocity head-on collisions between dust aggregates with various filling factors.
The aggregates used in their study were composed of $0.1~\si{\micro m}$-sized ice particles.
They found that the sticking probability is close to 100\% when the filling factor is lower than $0.45$ and the collision velocity is within the range of $1$--$10~\si{m.s^{-1}}$.
In contrast, the sticking probability is approximately 0\% when the filling factor is higher than $0.6$.
Similar results were also obtained by \citet{2013A&A...551A..65S}, although the aggregates used in their study were composed of $0.6~\si{\micro m}$-sized silicate particles.
Both \citet{2011ApJ...737...36W} and \citet{2013A&A...551A..65S} reported that dust aggregates with a filling factor $\le 0.4$ usually stick upon collision.

However, these numerical results are apparently inconsistent with the results from laboratory experiments.
\citet{2008ApJ...675..764L} performed microgravity experiments of collisions between millimeter- and centimeter-sized high-porosity dust aggregates of a filling factor of $0.15$.
They found that the sticking probability is approximately 60\% when the projectile mass is $0.1$--$10~\si{mg}$ and the collision velocity is $0.1$--$3~\si{m.s^{-1}}$.
\citet{2012Icar..218..688W} also performed microgravity experiments of collisions between two millimeter-sized aggregates of a filling factor of $0.35$.
They found that the sticking probability is close to 0\% when the collision velocity is higher than $10^{-2}~\si{m.s^{-1}}$ and the aggregate mass is larger than $0.1~\si{mg}$.
These experimental results seem to contradict the simulation results showing sticking at filling factors $\le 0.4$.
This discrepancy was already noted by earlier studies \citep[e.g.,][]{2011ApJ...737...36W, 2022MNRAS.509.5641S}.

Here, we propose that the discrepancy was due to the large difference in the size, i.e., the number of the constituent particles, of the used aggregates.
For example, the aggregates of \citet{2011ApJ...737...36W} consisted of only $10^{3}$--$10^{4}$ particles, whereas the aggregates of \citet{2008ApJ...675..764L} are likely to have contained $\sim 10^{9}$ particles assuming that their aggregates where $1~\si{mm}$ in radius.
We hypothesize that the threshold filling factor for collisional sticking/bouncing depends on aggregate size.
Indeed, some laboratory experiments already indicated that the sticking probability depends on aggregate size \citep[e.g.,][]{2008ApJ...675..764L, 2013Icar..225...75K}.

In this study, we test the hypothesis by numerical simulations using a soft-sphere discrete element method \citep[e.g.,][]{2022ApJ...933..144A, 2022ApJ...939..100A, 2023A&A...670L..21A}.
We succeed in reproducing the dependence of the sticking probability on aggregate size as indicated by laboratory experiments.
Our findings provide a hint to solve the long-standing problem of protoplanetary dust growth, that is, the discrepancy between numerical simulations and laboratory experiments for the bouncing barrier.

\section{Models}

We perform three-dimensional simulations of collisions between two equal-mass dust aggregates.
Our numerical code was originally developed by \citet{2007ApJ...661..320W} and is identical to that used in our previous studies \citep{2022ApJ...933..144A, 2022ApJ...939..100A, 2023A&A...670L..21A}.
In this study, we consider dust aggregates composed of spherical ice particles with a radius of $r_{1} = 0.1~\si{\micro m}$.
The material properties of ice are set to be equal to those assumed by \citet{2011ApJ...737...36W}; Young's modulus is $7~\si{GPa}$, Poisson's ratio is $0.25$, the surface energy is $100~\si{mJ.m^{-2}}$, and the material density is $1000~\si{kg.m^{-3}}$.
The critical rolling displacement is set to be $0.8~\si{nm}$.

We prepare spherical dust aggregates by the close-packing and particle-extraction (CPE) procedure \citep[e.g.,][]{2011ApJ...737...36W}.
We numerically produce cubic close-packed aggregates and randomly extract particles from the aggregates until their filling factors decrease to $\phi_{\rm agg} = 0.4$.
The orientation of the aggregates is randomly chosen.
The aggregate radius, $R_{\rm agg}$, is set to $R_{\rm agg} / r_{1} = 30$, $40$, $50$, $60$, and $70$.
The numbers of constituent particles in the target and projectile aggregates, $N_{\rm tar}$ and $N_{\rm pro}$, are given by $N_{\rm tar} = N_{\rm pro} = \phi_{\rm agg} {( R_{\rm agg} / r_{1} )}^{3}$.
The total number of particles in a simulation is therefore $N_{\rm tot} = N_{\rm tar} + N_{\rm pro} = 2 \phi_{\rm agg} {( R_{\rm agg} / r_{1} )}^{3}$.

We focus on head-on collisions.
The collision velocity of the two dust aggregates is set to $v_{\rm col} = 10^{( 0.1 i)}~\si{m.s^{-1}}$, where $i = 0$, $1$, ..., $10$. 
For each parameter set of ${( R_{\rm agg}, v_{\rm col} )}$, we perform four runs with different aggregates.
Thus, $4 \times 11 = 44$ runs have to be performed for each $R_{\rm agg}$, and we perform $44 \times 5 = 220$ runs in total.
The outcomes of all simulation runs are summarized in Appendix \ref{app:summary} (see Tables \ref{table1}--\ref{table5}).

The computational time for a single collision of two aggregates is approximately five weeks for $R_{\rm agg} / r_{1} = 70$ and $v_{\rm col} = 1~\si{m.s^{-1}}$, which is approximately proportional to ${R_{\rm agg}}^{4}$ and inversely proportional to $v_{\rm col}$.

\section{Results}

\begin{figure}
\centering
\includegraphics[width=0.95\columnwidth]{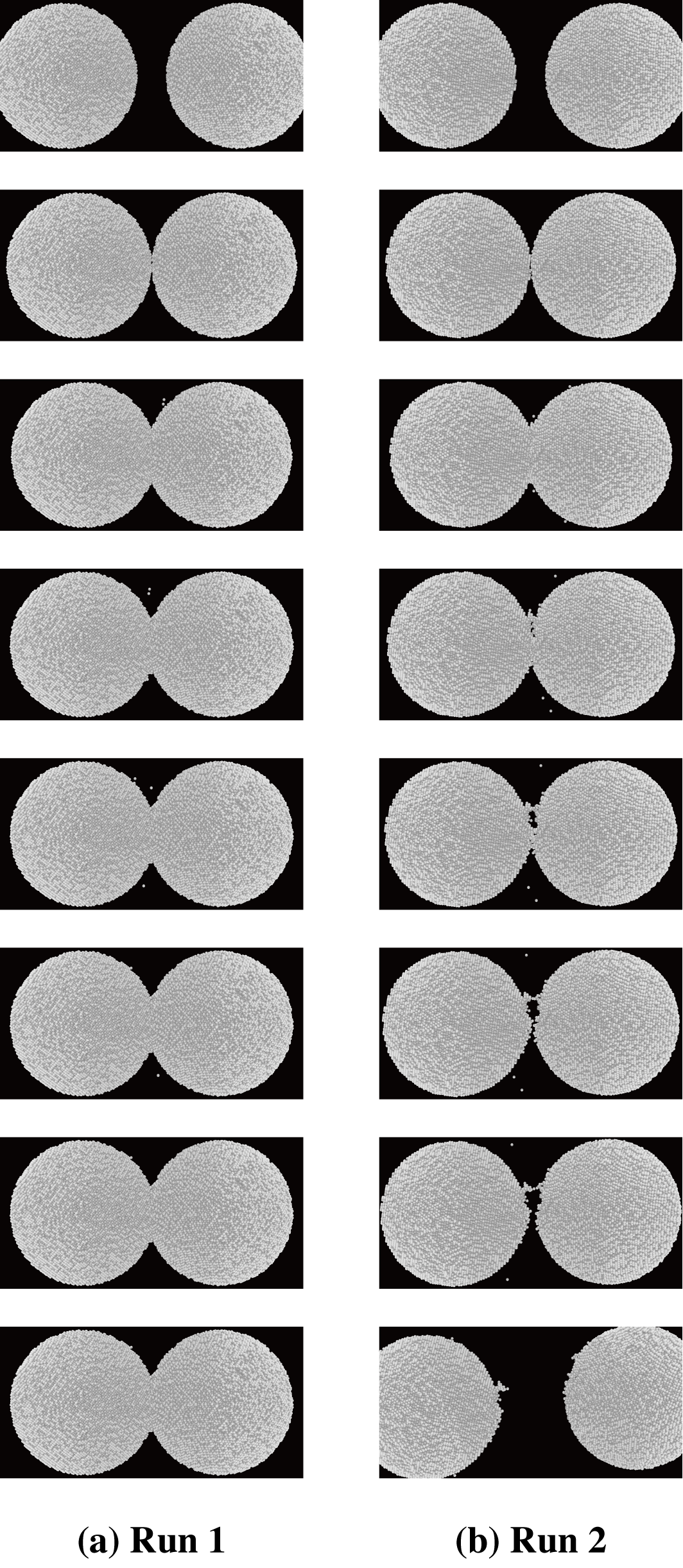}
\caption{
Snapshots of collisional outcomes.
Here, we set $R_{\rm agg} / r_{1} = 50$ (i.e., $N_{\rm tot} = 100000$) and $v_{\rm col} = 10~\si{m.s^{-1}}$.
The time interval is $0.2~\si{\micro s}$ from the first to the seventh snapshots, and the final snapshots were taken at $t = 4~\si{\micro s}$.
(a) Time series of snapshots for Run 1.
The number of constituent particles in the largest remnant was $N_{\rm lar} = 100000$ and we observed perfect sticking.
(b) Time series of snapshots for Run 2.
The number of constituent particles in the largest remnant was $N_{\rm lar} = 50256$ and we observed bouncing.
}
\label{fig.snapshot}
\end{figure}

Here, we show the results for $R_{\rm agg} / r_{1} = 50$ (i.e., $N_{\rm tot} = 100000$) and $v_{\rm col} = 10~\si{m.s^{-1}}$.
We find that the collisional outcomes strongly depend on the structure and orientation of the randomly prepared aggregates.
This is illustrated in Figure \ref{fig.snapshot}, where we show snapshots of two simulation runs (Runs 1 and 2, see Table \ref{table3}) with the same $R_{\rm agg} / r_{1} = 50$ (i.e., $N_{\rm tot} = 100000$) and $v_{\rm col} = 10~\si{m.s^{-1}}$ but with different aggregates.
Run 1 results in perfect sticking, whereas Run 2 results in bouncing.
This indicates that one must define the threshold between sticking and bouncing in a statistical manner, by averaging the collision outcomes over aggregate structures/orientations.

To quantify the collisional outcomes, we use the collisional growth efficiency, $f_{\rm gro}$, introduced by \citet{2013A&A...559A..62W}.
When the number of constituent particles in the largest remnant is $N_{\rm lar}$, $f_{\rm gro}$ is given by
\begin{equation}
f_{\rm gro} = \frac{N_{\rm lar} - N_{\rm tar}}{N_{\rm pro}}.
\end{equation}
This number is one for perfect sticking and zero for bouncing without mass transfer.

\begin{figure*}
\centering
\includegraphics[width=\columnwidth]{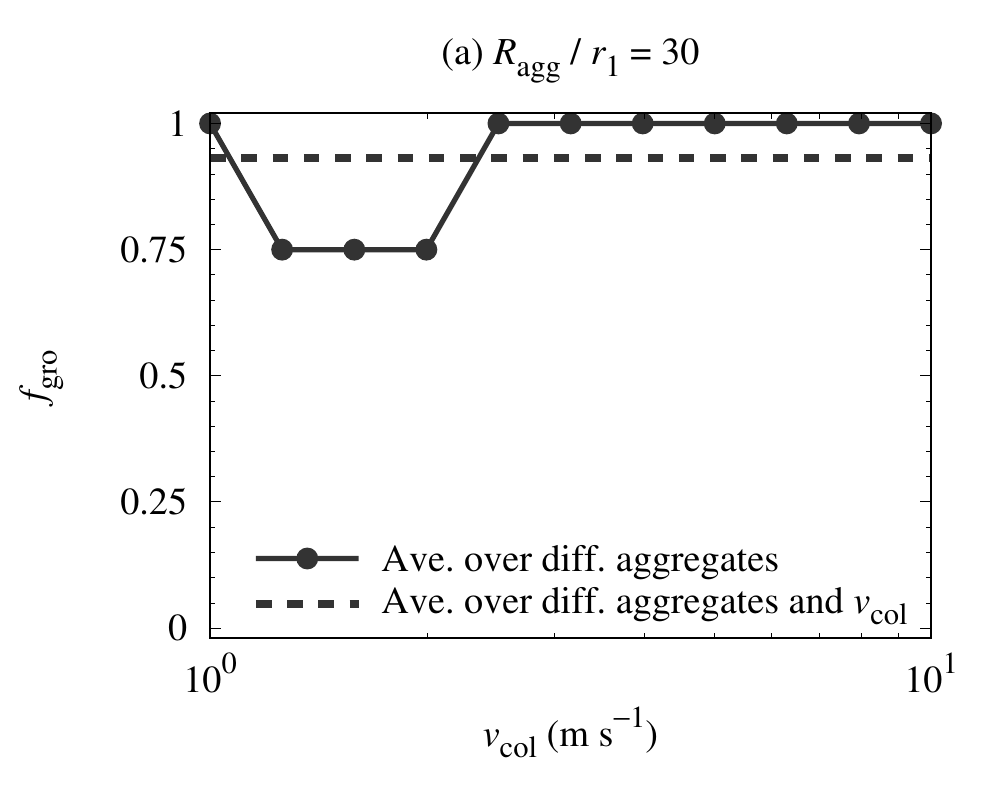}
\includegraphics[width=\columnwidth]{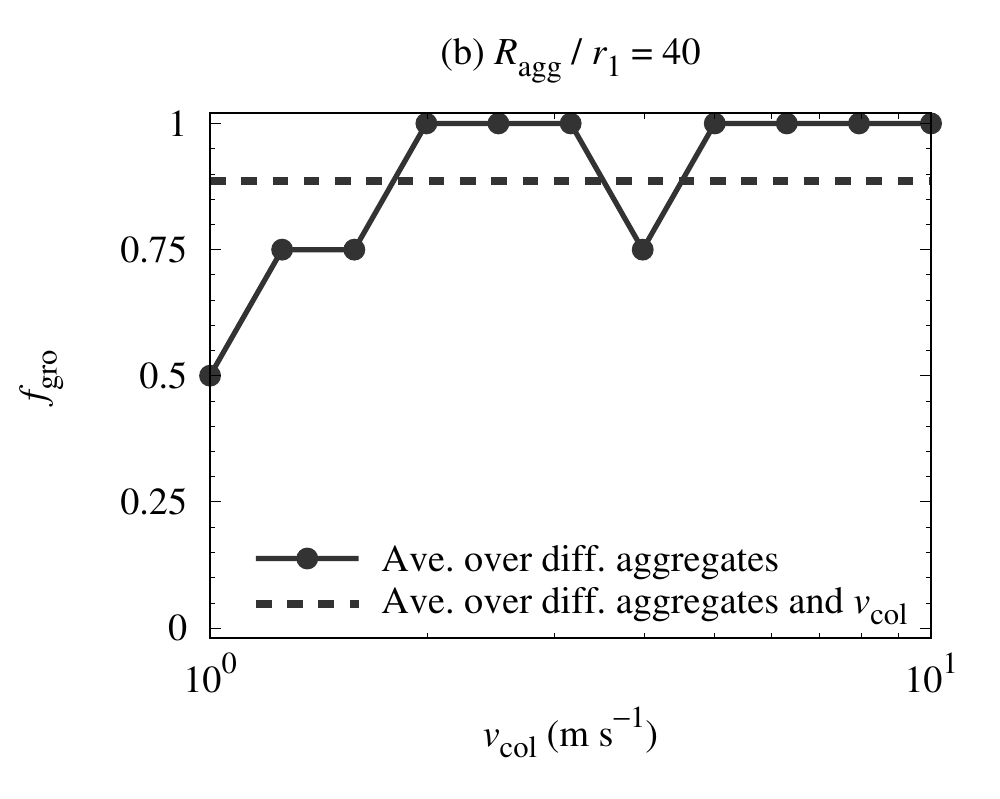}
\includegraphics[width=\columnwidth]{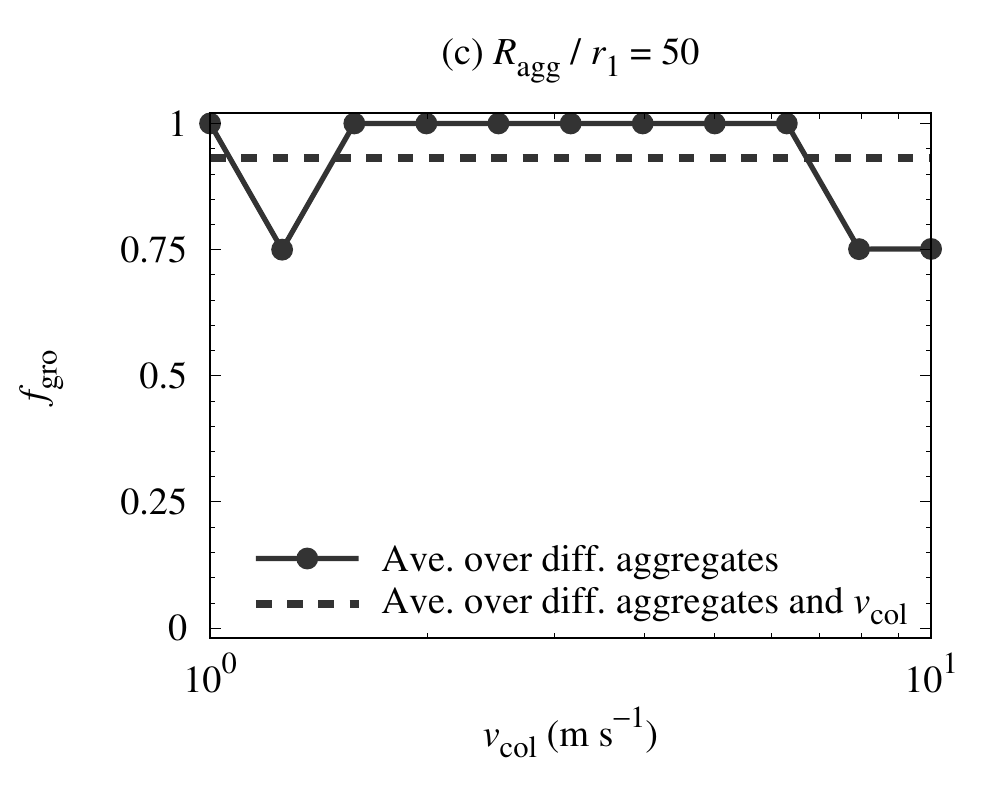}
\includegraphics[width=\columnwidth]{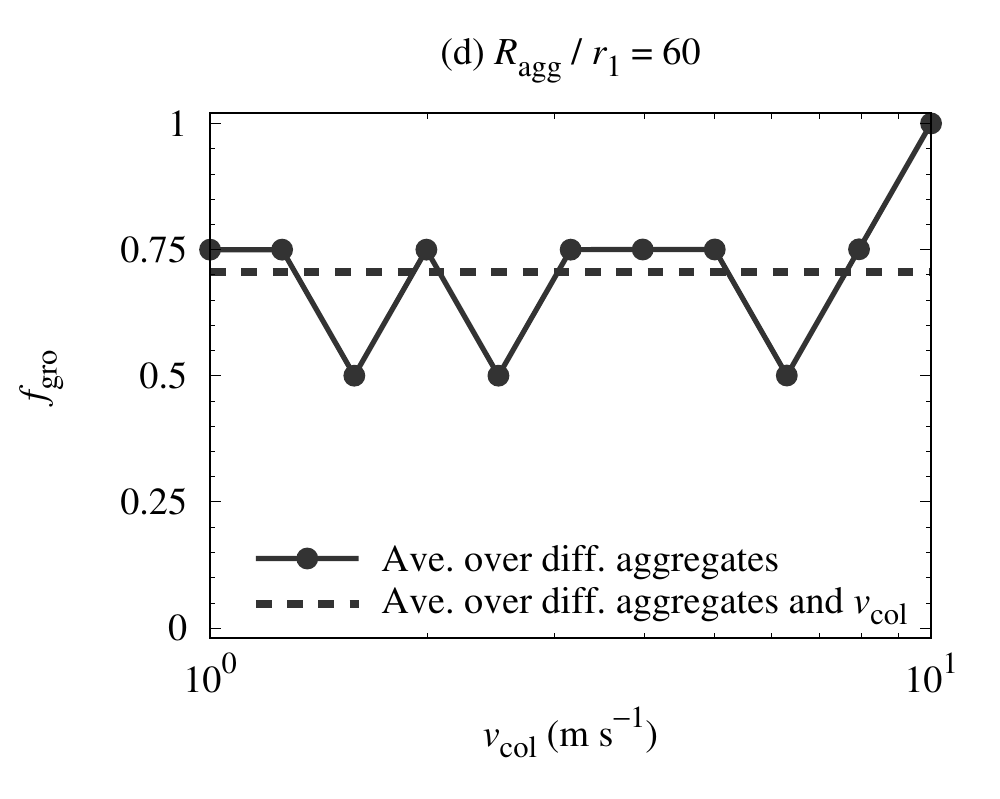}
\includegraphics[width=\columnwidth]{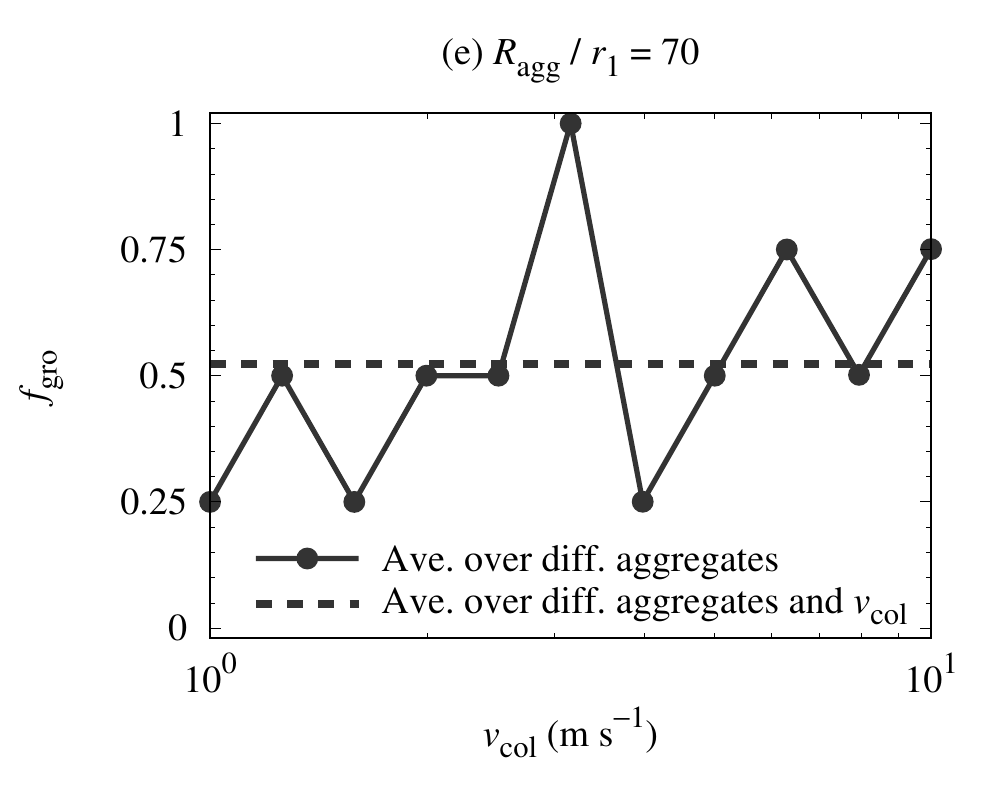}
\caption{
Averages of the collisional growth efficiency, $f_{\rm gro}$.
The solid lines represent the averages of $f_{\rm gro}$ over four different aggregates, whereas the dashed lines represent the averages over 44 different aggregates and $v_{\rm col}$.
(a) $R_{\rm agg} / r_{1} = 30$ ($N_{\rm tot} = 21600$).
(b) $R_{\rm agg} / r_{1} = 40$ ($N_{\rm tot} = 51200$).
(c) $R_{\rm agg} / r_{1} = 50$ ($N_{\rm tot} = 100000$).
(d) $R_{\rm agg} / r_{1} = 60$ ($N_{\rm tot} = 172800$).
(e) $R_{\rm agg} / r_{1} = 70$ ($N_{\rm tot} = 274400$).
}
\label{fig.fgro_R}
\end{figure*}

The solid lines in Figure \ref{fig.fgro_R} show the averages of $f_{\rm gro}$ over four different aggregates for different sets of ${( R_{\rm agg}, v_{\rm col} )}$.
The average of $f_{\rm gro}$ over the four runs takes a value of $\approx 0.25$, $0.5$, $0.75$, or $1$ because each collision results in either nearly perfect sticking ($f_{\rm gro} \approx 1$) or bouncing ($f_{\rm gro} \approx 0$).
Although the number of simulation runs for each parameter set is limited, no clear dependence of the average of $f_{\rm gro}$ on $v_{\rm col}$ is observed in the range of $1$--$10~\si{m.s^{-1}}$.
This feature is consistent with that observed in earlier studies \citep{2011ApJ...737...36W, 2013A&A...551A..65S}.
Thus we speculate that the average of $f_{\rm gro}$ is nearly independent of $v_{\rm col}$ in this velocity range.

Assuming that the aggregate average of $f_{\rm gro}$ is nearly independent of $v_{\rm col}$, we introduce the sticking probability, $f_{\rm gro, mean}$, as the average over different aggregates and $v_{\rm col}$ for each $R_{\rm agg}$.
The dashed lines in Figure \ref{fig.fgro_R} show $f_{\rm gro, mean}$ for each $R_{\rm agg}$.

\begin{figure}
\centering
\includegraphics[width=\columnwidth]{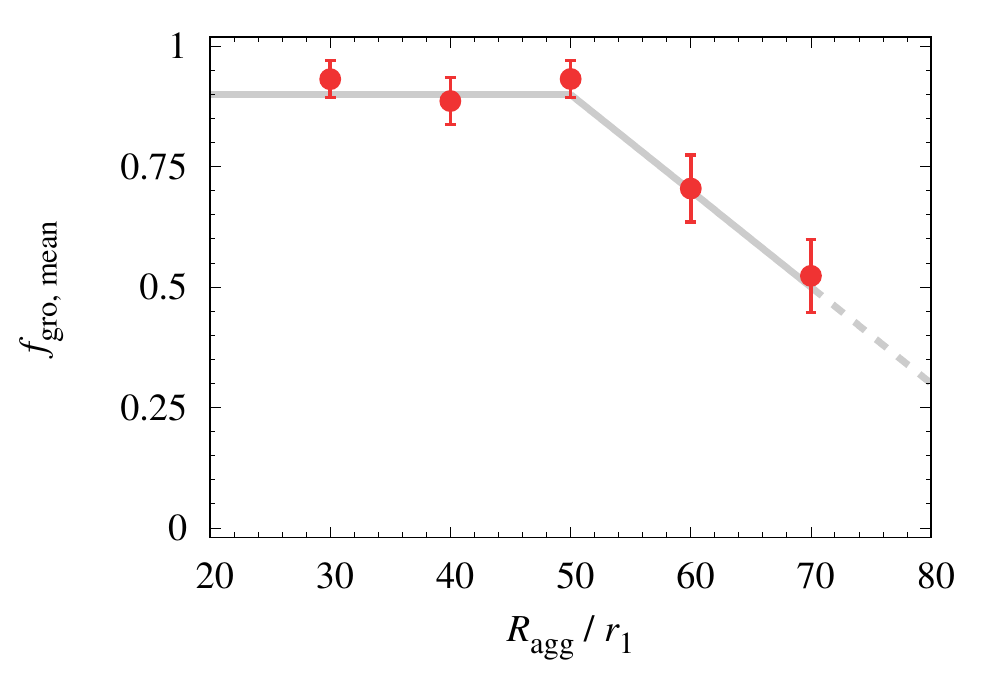}
\caption{
Sticking probability, $f_{\rm gro, mean}$, as a function of the aggregate radius, $R_{\rm agg}$.
The vertical error bars represent the standard error.
The gray line is a linear fit to guide the eye without theoretical interpretation.
}
\label{fig.fgro}
\end{figure}

Then we demonstrate that $f_{\rm gro, mean}$ decreases with increasing $R_{\rm agg}$.
Figure \ref{fig.fgro} shows $f_{\rm gro, mean}$ as a function of $R_{\rm agg}$.
We can see a clear dependence of $f_{\rm gro, mean}$ on $R_{\rm agg}$.
For $R_{\rm agg} / r_{1} \le 50$, $f_{\rm gro, mean}$ is close to $1$ and most of collisions result in sticking.
In contrast, for $R_{\rm agg} / r_{1} > 50$, $f_{\rm gro, mean}$ deviates from $1$ and decreases with increasing $R_{\rm agg}$.
Although the error bar is slightly large, $f_{\rm gro, mean}$ is approximately $0.5$ for $R_{\rm agg} / r_{1} = 70$, and $f_{\rm gro, mean}$ might be lower than $0.5$ when $R_{\rm agg} / r_{1}$ is larger than $70$.
In other words, the sticking probability decreases with increasing aggregate radius and falls below $0.5$ (and probably approaches to $0$) when colliding aggregates are sufficiently large.

We note that \citet{2013A&A...551A..65S} have reported similar results for dust aggregates made by CPE procedure.
They performed numerical simulations of collisions between two equal-mass dust aggregates for $R_{\rm agg} / r_{1} = 25$, $50$, and $83$.
They reported no clear difference in the sticking probability for aggregates with $R_{\rm agg} / r_{1} = 25$ and $50$.
However, they mentioned that the sticking probability for dust aggregates with $R_{\rm agg} / r_{1} = 83$ might be slightly lower than that for $R_{\rm agg} / r_{1} = 25$ and $50$.
Direct comparison between their results and ours is difficult because the material constants of constituent particles assumed in \citet{2013A&A...551A..65S} are different from those assumed in this study ($0.6~\si{\micro m}$-sized silicate particles and $0.1~\si{\micro m}$-sized ice particles, respectively).
As the material properties of constituent particles should affect the bouncing barrier \citep[e.g.,][]{2022MNRAS.509.5641S}, future studies on the bouncing barrier for dust aggregates with various settings for constituent particles would be necessary.

\section{Discussion}

We have shown that the sticking probability decreases with increasing aggregate radius (Figure \ref{fig.fgro}).
As discussed below, this can be interpreted as an indication of aggregate size dependence of the threshold filling factor for collisional sticking/bouncing.

\begin{figure}
\centering
\includegraphics[width=\columnwidth]{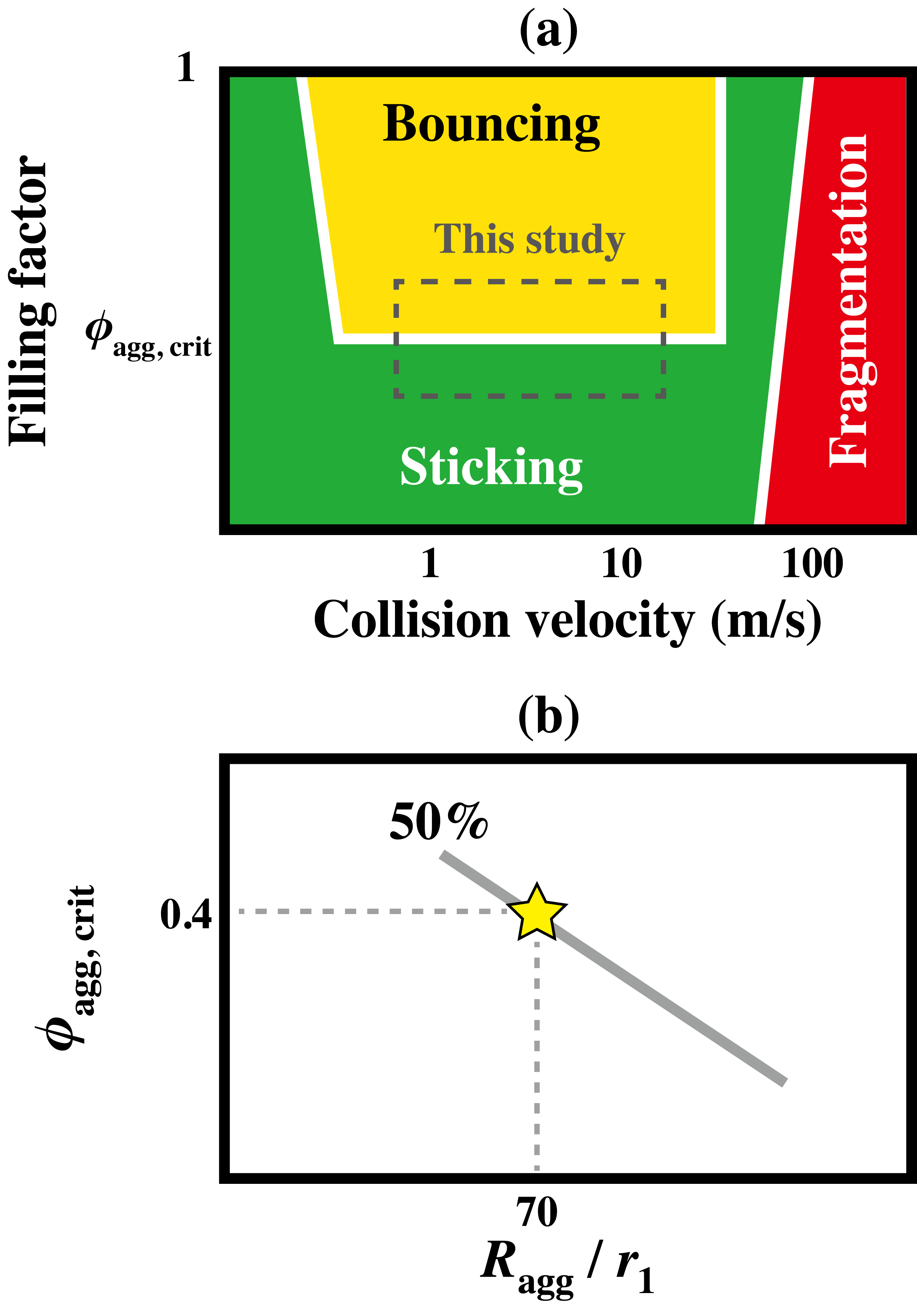}
\caption{
Map of the collisional outcomes.
(a) Collisional outcomes (sticking, bouncing, and fragmentation) as a function of the collision velocity and filling factor.
In the intermediate velocity range, two colliding aggregates bounce when their filling factor is higher than the threshold value, $\phi_{\rm agg, crit}$, and the collisional outcome is nearly independent of the collision velocity in this velocity range.
In this study, we feature the boundary between sticking and bouncing (i.e., the area enclosed by the dashed line). 
(b) Schematic of the dependence of $\phi_{\rm agg, crit}$ on $R_{\rm agg} / r_{1}$.
When we set the threshold sticking probability as 50\%, $\phi_{\rm agg, crit} \simeq 0.4$ when $R_{\rm agg} / r_{1} = 70$.
}
\label{fig.schematic}
\end{figure}

Figure \ref{fig.schematic} shows the map of the collisional outcomes.
In the low-velocity limit, both laboratory experiments \citep[e.g.,][]{2010A&A...513A..56G} and numerical simulations \citep[e.g.,][]{2011ApJ...737...36W} show that two colliding aggregates stick together.
In the intermediate velocity range, two colliding aggregates bounce when their filling factor is higher than the threshold value, $\phi_{\rm agg, crit}$.
Earlier numerical simulations \citep[e.g.,][]{2011ApJ...737...36W, 2013A&A...551A..65S} showed that the collisional outcome is nearly independent of the collision velocity in this velocity range, and we also confirmed this feature in Figure \ref{fig.fgro_R}.
In the high-velocity limit, the collisional outcome is fragmentation for both low- and high-density aggregates \citep[e.g.,][]{2009ApJ...702.1490W, 2013A&A...551A..65S}.
In addition, a narrow sticking window between bouncing and fragmentation regions had been reported in several numerical simulations \citep[e.g.,][]{2013A&A...551A..65S, OSINSKY2022127785}.

When we take the threshold sticking probability to be 50\%, the threshold filling factor for collisional sticking/bouncing is $\phi_{\rm agg, crit} \simeq 0.4$ for $R_{\rm agg} / r_{1} = 70$ (see Figure \ref{fig.schematic}(b)).
Simulations by \citet{2011ApJ...737...36W} show that the threshold filling factor for $R_{\rm agg} / r_{1} \simeq 18$ ranges between $0.45$ and $0.6$ for the same threshold sticking probability.
This indicates that $\phi_{\rm agg, crit}$ decreases with increasing $R_{\rm agg} / r_{1}$.

Experiments by \citet{2022MNRAS.509.5641S} also support our hypothesis that $\phi_{\rm agg, crit}$ is a decreasing function of $R_{\rm agg} / r_{1}$.
In laboratory experiments by \citet{2022MNRAS.509.5641S}, collisional bouncing was reported even if the filling factor is only $0.15$.
The size of dust aggregates used in \citet{2022MNRAS.509.5641S} was $R_{\rm agg} / r_{1} \gg 10^{3}$ and it is orders of magnitude larger than that for numerical simulations \citep[e.g.,][and this study]{2011ApJ...737...36W, 2013A&A...551A..65S}.
Their experimental results also support our hypothesis that $\phi_{\rm agg, crit}$ decreases with increasing $R_{\rm agg} / r_{1}$.
Future studies should test our hypothesis.

We note that $\phi_{\rm agg, crit}$ might also depend on $r_{1}$.
Recently, \citet{2022A&A...663A..57T} revealed that the results of optical and near-infrared polarimetric observations of protoplanetary disks are consistent with the presence of dust aggregates composed of submicron-sized particles \citep[see also][]{2023ApJ...944L..43T}.
As most of experimental studies used micron-sized particles instead of submicron-sized particles, understanding the dependence of $\phi_{\rm agg, crit}$ on the composition and size of constituent particles is essential to discuss the collisional behavior of aggregates in protoplanetary disks.
We will address this issue in future studies.

We found a strong dependence of the sticking probability on $R_{\rm agg} / r_{1}$ as shown in Figure \ref{fig.fgro}.
Although the reason of this strong dependence is still unclear, detailed analyses of temporal evolution of the packing structure of aggregates during collisions will provide a hint to unveil the reason.
\citet{2011ApJ...737...36W} reported that the energy dissipation occurs predominantly near the contact area of two colliding aggregates.
The geometry around the contact area and its temporal change during collisions should be studied in future detailed analyses.

\section{Conclusions}

Understanding the collisional behavior of dust aggregates is essential in the context of planet formation.
When the filling factor of colliding dust aggregates is higher than the threshold value, both numerical simulations and laboratory experiments have reported that low-velocity collisions of dust aggregates result in bouncing rather than sticking \citep[e.g.,][]{2008ApJ...675..764L, 2011ApJ...737...36W, 2013A&A...551A..65S}.
The threshold filling factor for sticking/bouncing was investigated in earlier studies, however, there is a large discrepancy between numerical and experimental results \citep[e.g.,][]{2011ApJ...737...36W, 2022MNRAS.509.5641S}.

One of the most notable differences between earlier numerical simulations and laboratory experiments is the number of constituent particles.
For example, in earlier numerical simulations by \citet{2011ApJ...737...36W}, the number of constituent particles is in the range of $10^{3}$--$10^{4}$, whereas the number of constituent particles in laboratory experiments by \citet{2008ApJ...675..764L} is on the order of $10^{9}$.

In this study, we hypothesized that the threshold filling factor for collisional sticking/bouncing would depend on the size of dust aggregates.
We tested this hypothesis by numerical simulations using soft-sphere discrete element method \citep[e.g.,][]{2022ApJ...933..144A, 2022ApJ...939..100A, 2023A&A...670L..21A}.
The filling factor of initial aggregates was set to $\phi_{\rm agg} = 0.4$, and we demonstrated that the sticking probability decreases with increasing the aggregate radius (Figure \ref{fig.fgro}).
Our results can be interpreted as the evidence that the threshold filling factor for collisional sticking/bouncing decreases with increasing the size of dust aggregates (Figure \ref{fig.schematic}).
Our findings give a qualitative answer for the apparent inconsistency between numerical simulations and laboratory experiments for the bouncing barrier; the large difference in the threshold filling factor would be due to not the artificial reasons but the difference in the size of dust aggregates in numerical simulations and laboratory experiments.

We acknowledge that the size range of dust aggregates investigated in this study is still limited and further investigations are needed in future studies.
We also fixed the filling factor of aggregates in this study, and we should investigate the collisional behavior of dust aggregates with different filling factors.
\citet{2011ApJ...737...36W} noted that the average coordination number might be the key parameter to understanding sticking/bouncing conditions instead of the filling factor.
As the average coordination number depends not only on the filling factor but also on the preparation procedure \citep[e.g.,][]{2013A&A...551A..65S, 2019PTEP.2019i3E02A}, we should perform numerical simulations using aggregates prepared by realistic procedures in future studies.
To perform numerical simulations with much larger aggregates and in wide parameter space, we need to address the acceleration of our numerical computations.
The calibration of material parameters by comparing numerical results with experimental outcomes would also be helpful in understanding the protoplanetary dust growth.


\section*{}

We thank the anonymous reveiwer for comments that greatly improved the paper.
Numerical computations were carried out on PC cluster at CfCA, NAOJ.
M.T.\ was supported by JSPS KAKENHI grant Nos.~JP22J00260 and JP22KJ1292.
H.T., E.K., and S.O.\ were supported by JSPS KAKENHI grant No.~JP18H05438.


%




\bibliography{sample631}{}
\bibliographystyle{aasjournal}

\clearpage
\appendix

\section{Summary of collisional outcomes}
\label{app:summary}

Tables \ref{table1}--\ref{table5} show the summary of $N_{\rm lar}$ for all simulations.
We performed 44 runs for each $R_{\rm agg} / r_{1}$ and calculated the mean value of $N_{\rm lar}$, $N_{\rm lar, mean}$.
The mean and the standard error of $N_{\rm lar}$ are obtained for each $R_{\rm agg} / r_{1}$: $N_{\rm lar, mean} = 20863.91 \pm 414.98$ for $R_{\rm agg} / r_{1} = 30$, $N_{\rm lar, mean} = 48291.30 \pm 1238.83$ for $R_{\rm agg} / r_{1} = 40$, $N_{\rm lar, mean} = 96601.68 \pm 1915.85$ for $R_{\rm agg} / r_{1} = 50$, $N_{\rm lar, mean} = 147287.05 \pm 6008.08$ for $R_{\rm agg} / r_{1} = 60$, and $N_{\rm lar, mean} = 208966.39 \pm  10442.86$ for $R_{\rm agg} / r_{1} = 70$.

\begin{table}
\caption{
Summary of $N_{\rm lar}$ for $R_{\rm agg} / r_{1} = 30$ ($N_{\rm tot} = 21600$).
}
\label{table1}
\centering
\begin{tabular}{crrrr}
{\bf $v_{\rm col}$}                   & {\bf Run 1} &  {\bf Run 2} &  {\bf Run 3} &  {\bf Run 4}   \\ \hline
$1.00~\si{m.s^{-1}}$                  & 21600       & 21600        & 21600        & 21600          \\
$1.26~\si{m.s^{-1}}$                  & 21600       & 21600        & 21600        & 10804          \\
$1.58~\si{m.s^{-1}}$                  & 21600       & 21600        & 21600        & 10804          \\
$2.00~\si{m.s^{-1}}$                  & 21600       & 21600        & 21600        & 10804          \\
$2.51~\si{m.s^{-1}}$                  & 21600       & 21600        & 21600        & 21600          \\
$3.16~\si{m.s^{-1}}$                  & 21600       & 21600        & 21600        & 21600          \\
$3.98~\si{m.s^{-1}}$                  & 21600       & 21600        & 21600        & 21600          \\
$5.01~\si{m.s^{-1}}$                  & 21600       & 21600        & 21600        & 21600          \\
$6.31~\si{m.s^{-1}}$                  & 21600       & 21600        & 21600        & 21600          \\
$7.94~\si{m.s^{-1}}$                  & 21600       & 21600        & 21600        & 21600          \\
$10.0~\si{m.s^{-1}}$                  & 21600       & 21600        & 21600        & 21600          \\ \hline
\end{tabular}
\end{table}

\begin{table}
\caption{
Summary of $N_{\rm lar}$ for $R_{\rm agg} / r_{1} = 40$ ($N_{\rm tot} = 51200$).
}
\label{table2}
\centering
\begin{tabular}{crrrr}
{\bf $v_{\rm col}$}                   & {\bf Run 1} &  {\bf Run 2} &  {\bf Run 3} &  {\bf Run 4}   \\ \hline
$1.00~\si{m.s^{-1}}$                  & 51200       & 25601        & 51200        & 25606          \\
$1.26~\si{m.s^{-1}}$                  & 51200       & 51200        & 51200        & 25606          \\
$1.58~\si{m.s^{-1}}$                  & 51200       & 51200        & 51200        & 25603          \\
$2.00~\si{m.s^{-1}}$                  & 51200       & 51200        & 51200        & 51200          \\
$2.51~\si{m.s^{-1}}$                  & 51200       & 51200        & 51200        & 51200          \\
$3.16~\si{m.s^{-1}}$                  & 51200       & 51200        & 51200        & 51200          \\
$3.98~\si{m.s^{-1}}$                  & 51200       & 25601        & 51200        & 51200          \\
$5.01~\si{m.s^{-1}}$                  & 51200       & 51200        & 51200        & 51200          \\
$6.31~\si{m.s^{-1}}$                  & 51200       & 51200        & 51200        & 51200          \\
$7.94~\si{m.s^{-1}}$                  & 51200       & 51200        & 51200        & 51200          \\
$10.0~\si{m.s^{-1}}$                  & 51200       & 51200        & 51200        & 51200          \\ \hline
\end{tabular}
\end{table}

\begin{table}
\caption{
Summary of $N_{\rm lar}$ for $R_{\rm agg} / r_{1} = 50$ ($N_{\rm tot} = 100000$).
}
\label{table3}
\centering
\begin{tabular}{crrrr}
{\bf $v_{\rm col}$}                   & {\bf Run 1} &  {\bf Run 2} &  {\bf Run 3} &  {\bf Run 4}   \\ \hline
$1.00~\si{m.s^{-1}}$                  & 100000      & 100000       & 100000       & 100000         \\
$1.26~\si{m.s^{-1}}$                  & 100000      & 50001        & 100000       & 100000         \\
$1.58~\si{m.s^{-1}}$                  & 100000      & 100000       & 100000       & 100000         \\
$2.00~\si{m.s^{-1}}$                  & 100000      & 100000       & 100000       & 100000         \\
$2.51~\si{m.s^{-1}}$                  & 100000      & 100000       & 100000       & 100000         \\
$3.16~\si{m.s^{-1}}$                  & 100000      & 100000       & 100000       & 100000         \\
$3.98~\si{m.s^{-1}}$                  & 100000      & 100000       & 100000       & 100000         \\
$5.01~\si{m.s^{-1}}$                  & 100000      & 100000       & 100000       & 100000         \\
$6.31~\si{m.s^{-1}}$                  & 100000      & 100000       & 100000       & 100000         \\
$7.94~\si{m.s^{-1}}$                  & 100000      & 50217        & 100000       & 100000         \\
$10.0~\si{m.s^{-1}}$                  & 100000      & 50256        & 100000       & 100000         \\ \hline
\end{tabular}
\end{table}

\begin{table}
\caption{
Summary of $N_{\rm lar}$ for $R_{\rm agg} / r_{1} = 60$ ($N_{\rm tot} = 172800$).
}
\label{table4}
\centering
\begin{tabular}{crrrr}
{\bf $v_{\rm col}$}                   & {\bf Run 1} &  {\bf Run 2} &  {\bf Run 3} &  {\bf Run 4}   \\ \hline
$1.00~\si{m.s^{-1}}$                  & 86400       & 172800       & 172800       & 172800         \\
$1.26~\si{m.s^{-1}}$                  & 86404       & 172800       & 172800       & 172800         \\
$1.58~\si{m.s^{-1}}$                  & 86404       & 86402        & 172800       & 172800         \\
$2.00~\si{m.s^{-1}}$                  & 86408       & 172800       & 172800       & 172800         \\
$2.51~\si{m.s^{-1}}$                  & 172800      & 86413        & 172800       & 86407          \\
$3.16~\si{m.s^{-1}}$                  & 172800      & 86415        & 172800       & 172800         \\
$3.98~\si{m.s^{-1}}$                  & 172800      & 172800       & 86498        & 172800         \\
$5.01~\si{m.s^{-1}}$                  & 172800      & 86445        & 172800       & 172800         \\
$6.31~\si{m.s^{-1}}$                  & 172800      & 86549        & 172800       & 86482          \\
$7.94~\si{m.s^{-1}}$                  & 172800      & 172800       & 86603        & 172800         \\
$10.0~\si{m.s^{-1}}$                  & 172800      & 172800       & 172800       & 172800         \\ \hline
\end{tabular}
\end{table}

\begin{table}
\caption{
Summary of $N_{\rm lar}$ for $R_{\rm agg} / r_{1} = 70$ ($N_{\rm tot} = 274400$).
}
\label{table5}
\centering
\begin{tabular}{crrrr}
{\bf $v_{\rm col}$}                   & {\bf Run 1} &  {\bf Run 2} &  {\bf Run 3} &  {\bf Run 4}   \\ \hline
$1.00~\si{m.s^{-1}}$                  & 137200      & 137202       & 274400       & 137212         \\
$1.26~\si{m.s^{-1}}$                  & 137201      & 137205       & 274400       & 274398         \\
$1.58~\si{m.s^{-1}}$                  & 137204      & 137211       & 274400       & 137207         \\
$2.00~\si{m.s^{-1}}$                  & 274400      & 274400       & 137202       & 137203         \\
$2.51~\si{m.s^{-1}}$                  & 137235      & 274400       & 137214       & 274398         \\
$3.16~\si{m.s^{-1}}$                  & 274400      & 274400       & 274400       & 274398         \\
$3.98~\si{m.s^{-1}}$                  & 137363      & 137223       & 274400       & 137298         \\
$5.01~\si{m.s^{-1}}$                  & 137332      & 274400       & 137207       & 274398         \\
$6.31~\si{m.s^{-1}}$                  & 274400      & 274400       & 274400       & 137351         \\
$7.94~\si{m.s^{-1}}$                  & 274400      & 274400       & 137762       & 137616         \\
$10.0~\si{m.s^{-1}}$                  & 274400      & 274400       & 137681       & 274400         \\ \hline
\end{tabular}
\end{table}




\end{document}